\documentclass[journal=jacsat,manuscript=article]{achemso}

\usepackage[version=3]{mhchem} 
\usepackage{siunitx}
\usepackage{microtype}
\usepackage[capitalize,noabbrev]{cleveref}

\usepackage{xcolor}

\author{Frederico B. Sousa\footnotemark[1]\footnotetext{$^\ast$ F.B.S. and B.Z. contributed equally to this paper as regards experimental and computational aspects respectively.}}
\affiliation{Departamento de F\'isica, Universidade Federal de Minas Gerais, Belo Horizonte, Minas Gerais 30123-970, Brazil}

\author{Boyang Zheng\footnotemark[1]}
\affiliation{Department of Physics, The Pennsylvania State University, University Park, PA 16802, United States of America}
\alsoaffiliation{The Materials Research Institute, The Pennsylvania State University, University Park, PA 16802, United States of America}

\author{Mingzu Liu}
\affiliation{Department of Physics, The Pennsylvania State University, University Park, PA 16802, United States of America}
\alsoaffiliation{Center for 2-Dimensional and Layered Materials, The Pennsylvania State University, University Park, PA 16802, United States of America}

\author{Geovani C. Resende}
\affiliation{Departamento de F\'isica, Universidade Federal de Minas Gerais, Belo Horizonte, Minas Gerais 30123-970, Brazil}

\author{Da Zhou}
\affiliation{Department of Physics, The Pennsylvania State University, University Park, PA 16802, United States of America}
\alsoaffiliation{Center for 2-Dimensional and Layered Materials, The Pennsylvania State University, University Park, PA 16802, United States of America}

\author{Marcos A. Pimenta}
\affiliation{Departamento de F\'isica, Universidade Federal de Minas Gerais, Belo Horizonte, Minas Gerais 30123-970, Brazil}
\alsoaffiliation{Departamento de F\'isica, Universidade Federal de Ouro Preto, Ouro Preto, Minas Gerais 35400-000, Brazil}

\author{Mauricio Terrones}
\affiliation{Center for 2-Dimensional and Layered Materials, The Pennsylvania State University, University Park, PA 16802, United States of America}
\alsoaffiliation{Department of Physics, The Pennsylvania State University, University Park, PA 16802, United States of America}
\alsoaffiliation{Department of Materials Science and Engineering, The Pennsylvania State University, University Park, PA 16802, United States of America}

\author{Vincent H. Crespi}
\affiliation{Department of Physics, The Pennsylvania State University, University Park, PA 16802, United States of America}
\alsoaffiliation{The Materials Research Institute, The Pennsylvania State University, University Park, PA 16802, United States of America}
\alsoaffiliation{Department of Materials Science and Engineering, The Pennsylvania State University, University Park, PA 16802, United States of America}
\alsoaffiliation{Center for 2-Dimensional and Layered Materials, The Pennsylvania State University, University Park, PA 16802, United States of America}

\author{Leandro M. Malard}
\affiliation{Departamento de F\'isica, Universidade Federal de Minas Gerais, Belo Horizonte, Minas Gerais 30123-970, Brazil}
\email{lmalard@fisica.ufmg.br}

\title[Vanadium Doped WS2]
{Effects of Vanadium Doping on the Optical Response and Electronic Structure of WS$_{2}$ Monolayers}
    
\abbreviations{IR,NMR,UV}

\begin{document}



\newpage
\begin{abstract}

Two-dimensional dilute magnetic semiconductors has been recently reported in semiconducting transition metal dichalcogenides by the introduction of spin-polarized transition metal atoms as dopants. This is the case of vanadium-doped WS$_2$ and WSe$_2$ monolayers, which exhibits a ferromagnetic ordering even above room temperature. However, a broadband characterization of their electronic band structure and its dependence on vanadium concentration is still lacking. Therefore, here we perform power-dependent photoluminescence, resonant four-wave mixing, and differential reflectance spectroscopy to study the optical transitions close to the A exciton energy of vanadium-doped WS$_2$ monolayers with distinct concentrations. Instead of a single A exciton peak, vanadium-doped samples exhibit two photoluminescence peaks associated with transitions to occupied and unoccupied bands. 
Moreover, resonant Raman spectroscopy and resonant second-harmonic generation measurements revealed a blueshift in the B exciton but no energy change in the C exciton as vanadium is introduced in the monolayers. Density functional theory calculations showed that the band structure is sensitive to the Hubbard \(U\) correction for vanadium and several scenarios are  proposed to explain the two photoluminescence peaks around the A exciton energy region. Our work provides the first broadband optical characterization of these two-dimensional dilute magnetic semiconductors, shedding light on the novel electronic features of WS$_{2}$ monolayers which are tunable by the vanadium concentration. 


\end{abstract}

\section{Introduction}

Tuning the electronic, optical, magnetic, and/or physico-chemical properties of two-dimensional (2D) transition metal dichalcogenides (TMDs) through defect engineering provides multiple pathways for diverse applications.\cite{Lin2016,Hong2017,Wang2018,Liang2021,Ippolito2022}
For instance, a dilute magnetic semiconductor (DMS) \cite{Dietl2014} can be achieved by incorporating transition metal atoms as substitutional defects in TMD monolayers \cite{Ramasubramaniam2013,Wang2016,Habib2018,Yun2020,Pham2020,Zhang2020}. Vanadium-doped WSe$_{2}$ \cite{Yun2020,Pham2020} and WS$_{2}$ \cite{Zhang2020} samples have shown a long-range ferromagnetic ordering even above room temperature, thus opening  possibilities for spintronic devices fabrication. However, among the significant challenges to be overcome in this field such as the gate-tunability manifestation and the necessity of enhanced magnetic moments in these materials \cite{Lee2023,OrtizJimenez2023}, there is a fundamental need for broadband optical and electronic structure characterization of these doped 2D semiconductors.

Here we study pristine and V-doped WS$_{2}$ monolayer samples with three different atomic percentages (at\%) of vanadium to explore the dependence of the optical response on the vanadium concentration near the three main excitonic energies (A, B, and C excitons), using power-dependent photoluminescence (PL), resonant Raman spectroscopy, resonant four-wave mixing (FWM) and second-harmonic generation (SHG) spectroscopies, and differential reflectance spectroscopy. The power-dependent PL, resonant FWM, and differential reflectance spectroscopy showed a splitting of the A exciton with one peak blueshifting and the other one redshifting under increasing vanadium doping. The FWM measurement further showed that the lower-energy peak is associated with a transition to an unoccupied electronic band, proving the existence of free holes which were proposed by other groups~\cite{Duong2019,Duong2020,Song2021} to mediate a Zener-type magnetic exchange, while the higher-energy peak reveals a transition to an occupied band. The Raman resonances show a blueshifting B exciton energy, while SHG resonant profiles show no change in the C exciton energy with increasing vanadium concentration. 
Density functional theory (DFT) calculations for the band structure of V-doped WS$_{2}$ with a Hubbard \(U\) correction (\(U = \qtyrange{0}{5}{eV}\)) for vanadium, including transition dipole moments, were performed to elucidate the optical transitions observed in the experiments.  Our work also shows the potential of resonant Raman and nonlinear techniques to probe new features in the electronic band structure of 2D semiconductors.

\section{Results and Discussion}
\subsection{Power-dependent photoluminescence in V-doped WS$_2$}

\begin{figure}[!htb]
 \centering
 \includegraphics[scale=0.4]{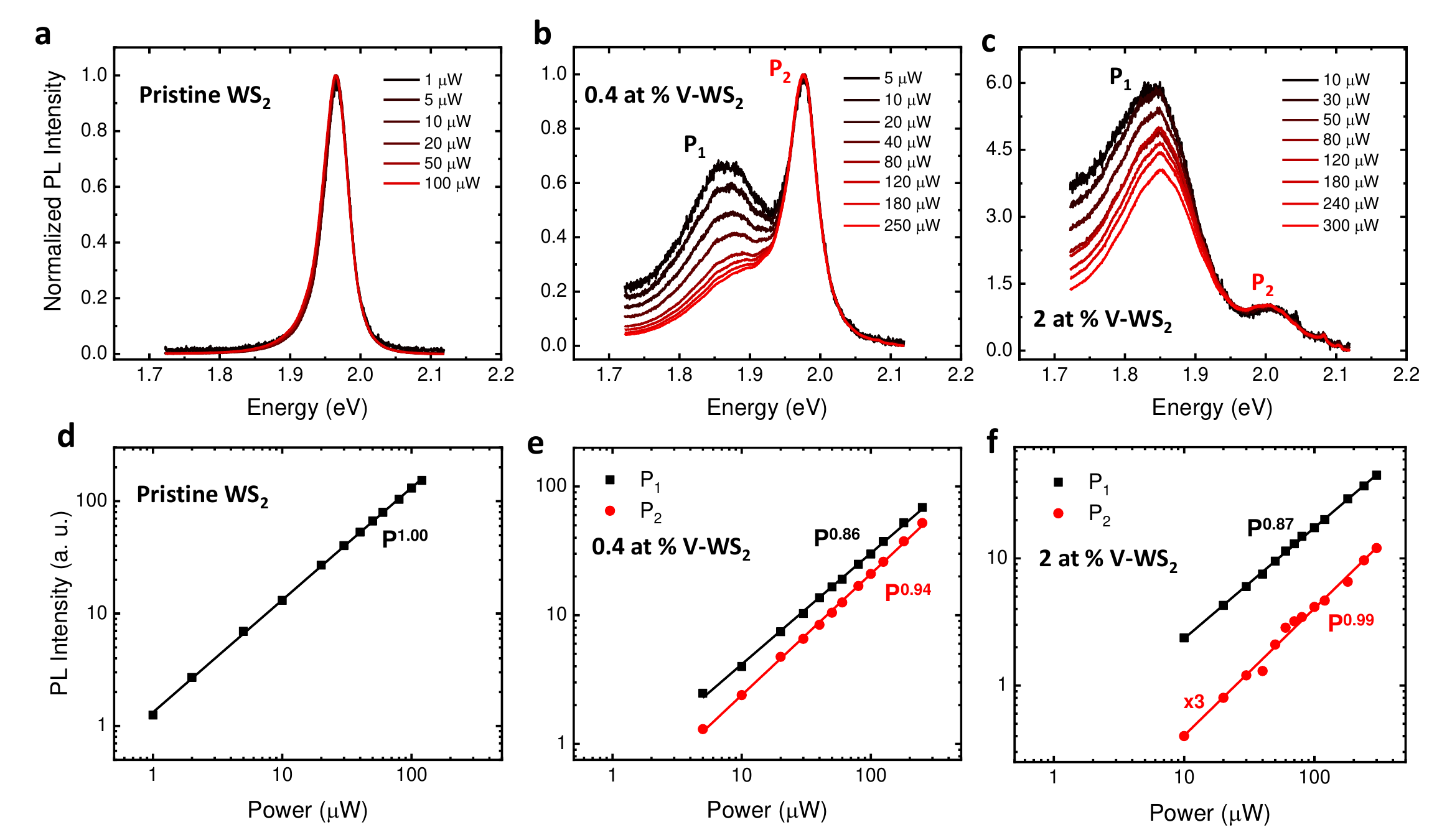} 
 \caption{{\small {\bf a-c} Normalized PL spectra for a pristine ({\bf a}) and 0.4 at\% ({\bf b}) and 2.0 at\% ({\bf c}) vanadium doped WS$_{2}$ monolayers for distinct pump powers. All spectra were taken with an excitation wavelength of 561 nm and are normalized by the higher energy peak maximum. {\bf d-f} Power-dependent PL intensity showing a linear power dependence for the pristine WS$_{2}$ peak and V-doped WS$_{2}$ higher energy peaks (P$_{2}$) and a sublinear dependence for the V-doped WS$_{2}$ lower energy peaks (P$_{1}$).
 }}
 \label{figPL}
 \end{figure}

The pristine and V-doped WS$_{2}$ monolayers used in this work were synthesized by single-step chemical vapor deposition (CVD) as described by Zhang {\it et al.} \cite{Zhang2020} (a detailed description of the samples growth can be found in Methods). To confirm the presence of substitutional V atoms at W sites and to determine their concentration, high-angular annular dark-field scanning transmission electron microscopy (HAADF-STEM) imaging was performed in 4 distinct samples as shown in Supporting Figures S1a-d. Besides the pristine sample (Figure S1a), a statistical analysis of Figures S1b-d revealed vanadium concentrations of 0.4 at\%, 2.0 at\%, and 8.0 at\%, respectively. In order to experimentally study the electronic band structure modifications in these samples due to vanadium doping, we employed power-dependent PL measurements with a 561 nm laser excitation energy as shown in the normalized spectra of Figures \ref{figPL}a-c (PL measurements for the 8.0 at\% sample are shown in Supporting Figure S5). The normalization of the PL spectra was performed with respect to the higher-energy PL peak (from the A exciton in the case of pristine WS$_{2}$). While the pristine sample exhibits a single PL peak at 1.96 eV related to A exciton emission,\cite{Zeng2013,McCreary2016} the doped samples display one peak in the energy range of 1.8--1.9 eV (we call P$_{1}$) and a second peak around 2.0 eV (P$_{2}$), as reported by Zhang {\it et al.} \cite{Zhang2023}. It can be noted that after increasing vanadium doping the lower energy peak P$_{1}$ redshifts --- in agreement with the band gap reduction predicted from previous DFT calculations \cite{Zhang2020}. In addition, the higher-energy PL peak P$_{2}$ blueshifts. Moreover, the substitutional defects broadened the PL peak and quenched the integrated PL intensity, similar to other works.\cite{Zhang2019,Li2021,Rosa2022} All PL spectra were fitted by two Gaussian peaks and their intensities as a function of laser power are shown in Figures \ref{figPL}d-f. Beside the expected linear power dependence of the A exciton peak in pristine WS$_{2}$, there is a linear (for P$_{2}$) and sublinear (for P$_{1}$) power dependencies for V-doped samples (also shown in Supporting Figures S2-S5). 
According to Refs. \cite{Schmidt1992, Spindler2019}, the power-dependent PL measurements reveal that the P$_{2}$ peak is associated with an exciton formation and the P$_{1}$ peak is related to a radiative recombination process from a donor or to an acceptor level. However, previous DFT calculations showed no mid gap states for V-doped WS$_{2}$ monolayers \cite{Zhang2020}, while HAADF-STEM data present no indication of higher S mono-vacancy defects for the V-doped samples. Therefore, this sublinear power dependence might be related to an acceptor state instead of a donor defective level. In order to gain more experimental evidence about the nature of these new electronic states, we next perform FWM and differential reflectance spectroscopy experiments.

\subsection{Resonant four-wave mixing in V-doped WS$_2$}

\begin{figure}[!htb]
 \centering
 \includegraphics[scale=0.4]{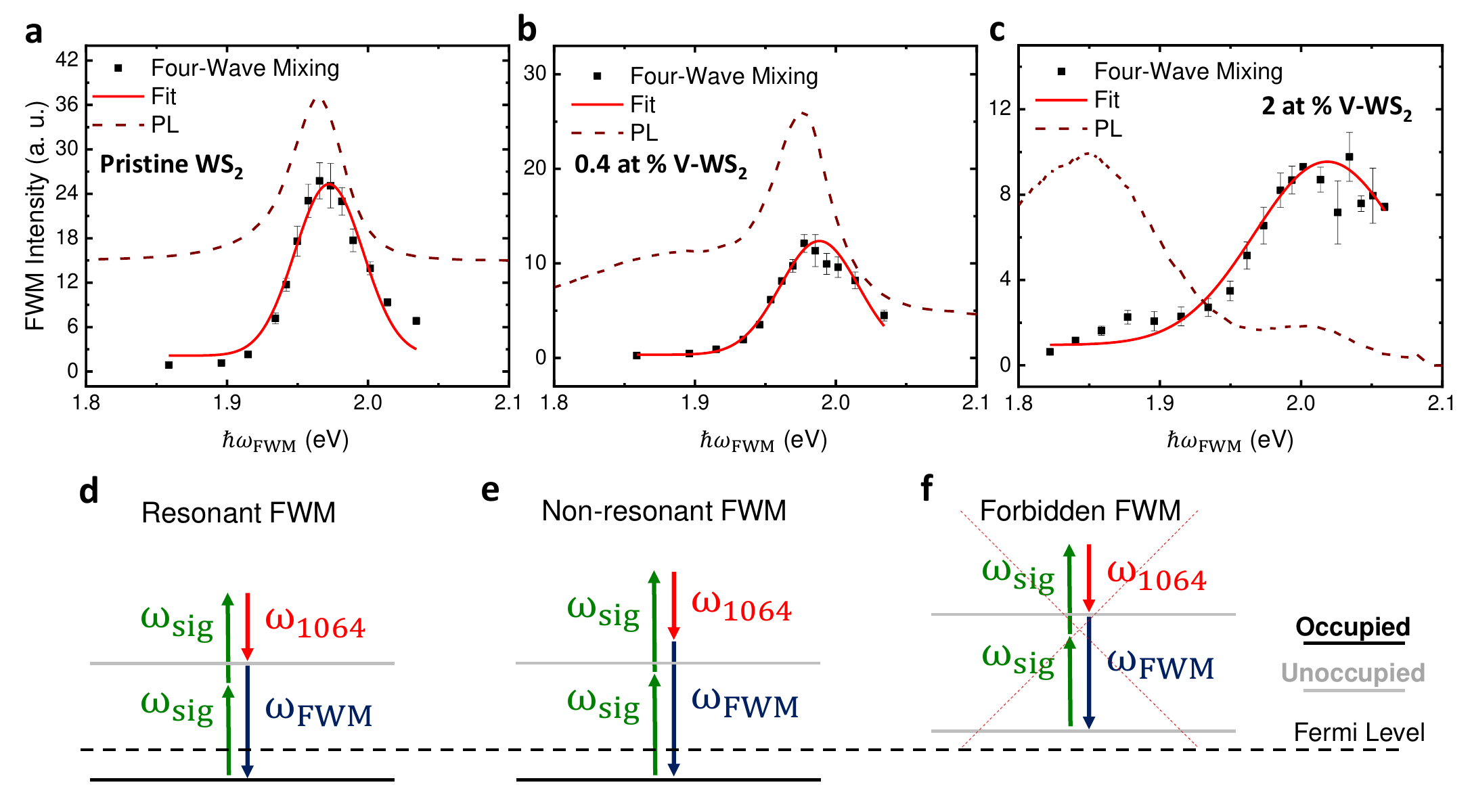} 
 \caption{{\small {\bf a-c} FWM resonant profile (black squares) and its Gaussian fits (red line) together with the PL spectra (dashed brown line) for the pristine ({\bf a}), 0.4 at\% ({\bf b}) and 2.0 at\% ({\bf c}) vanadium doped WS$_{2}$ monolayers. The FWM resonances align with the higher energy PL peak for the V-doped samples. {\bf d-f} FWM energy diagrams for resonant ({\bf d}), non-resonant ({\bf e}), and forbidden ({\bf f}) conditions. The colored arrows correspond to the photons of the FWM process as described by Equation \ref{eq:FWM}. The black lines are associated with occupied states, the gray lines with unoccupied states and the dashed line with the Fermi level.
 }}
 \label{figFWM}
 \end{figure}

As shown by Lafeta {\it et al.},\cite{Lafeta2021} nonlinear optical techniques such as FWM can be used to determine excitonic energies (Figure~\ref{figFWM}). FWM is a third-order nonlinear optical process in which three photons interact in a nonlinear material to generate a fourth photon at a linear combination of the three incident photon frequencies (with coefficients $\pm$1). Here we measured a degenerate FWM in which two incident photons with the same variable frequency $\omega_{\mathrm{pump}}$ (from 720 to 950 nm) interact with a third photon of a fixed frequency $\omega_{1064}$ (1064 nm) to generate a fourth photon with frequency:
\begin{equation}
    \omega_{\mathrm{FWM}} = 2\omega_{\mathrm{pump}} - \omega_{1064}.
    \label{eq:FWM}
\end{equation}
The wide wavelength range of the resulting $\omega_{FWM}$ (544 to 858 nm) allowed us to probe resonant responses over the energy window of the WS$_{2}$ monolayer A exciton (600 to 680 nm).  Figures \ref{figFWM}a-c show the PL spectra and the FWM resonant profiles with their Gaussian fits for three samples (pristine and doped \ce{WS2} with 0.4 at\% and 2.0 at\% vanadium). The resonant behavior of the FWM measurement for the pristine WS$_{2}$ monolayer (Figure~\ref{figFWM}a) is in good agreement with its single PL peak. The few meV energy shift between the FWM resonant response and the PL peak is presumably due to a Stokes shift.\cite{Zhao2013,Fan2019}
Figures \ref{figFWM}b,c show that both 0.4 at\% and 2.0 at\% vanadium-doped WS$_{2}$ monolayers present a strong FWM resonant response at the energy of the P$_{2}$ PL peak, as well as the same Stokes shift. The 0.4 at\% doped sample shows no resonance close to the lower-energy P$_{1}$ PL peak, while the 2.0 at\% doped sample displays a weak FWM resonant feature around the P$_{1}$ PL peak energy. As shown by Equation~\ref{eq:FWM} and Figures~\ref{figFWM}d-f, the FWM process requires energy conservation as it brings the electron back to the same occupied electronic band from which it was excited, meaning that the FWM allowed processes have final states below the Fermi level. Therefore, the final state of the P$_2$ peak is occupied, while the absence of a relevant FWM resonance close to the P$_1$ peak energy suggests a lower occupancy level of its final state. Since vanadium doping introduces free holes in TMD monolayers \cite{Zhang2020,Duong2019,Duong2020}, the final electronic state associated with the P$_{1}$ peak could be these itinerant holes. The FWM results for WS$_{2}$ monolayers at 8 at\% vanadium concentration are not shown due to its weak intensity.
In addition, differential reflectance measurements presented in Supporting Figure S6 --- which also probe electronic excitations only from occupied bands --- only clearly show the P$_2$ peak for doped samples, in agreement with the results described above.
 
\subsection{Resonant Raman scattering in V-doped WS$_2$}

In order to further characterize the electronic band structure modifications in the monolayer WS$_{2}$ due to vanadium doping, we performed resonant Raman spectroscopy measurements to probe the B exciton energy range. Supporting Figure S7 shows Raman spectra of these pristine and V-doped WS$_{2}$ monolayers for 14 different laser lines in the 250--550 cm$^{-1}$ spectral range (low-frequency Raman spectra were also measured and are shown in Supporting Figure S8). All spectra were normalized by the silicon peak intensity (520 cm$^{-1}$ peak), considering its Raman cross-section \cite{Lautenschlager1987} for each laser line. A strong dependence of the Raman peaks intensities on the excitation energy can be observed. This dependence, plotted as Raman excitation profile (REP), varies with the doping concentration, indicating different resonant excitation behaviors among samples; it is therefore useful to find the B exciton energy. The intensity of a first-order Raman mode is~\cite{Jorio2011}
\begin{equation}
    I(E_{\mathrm{pump}}) = C \left | \sum_{m,n} \frac{\langle f \vert H_{e-r} \vert n \rangle \langle n \vert H_{e-ph} \vert m \rangle \langle m \vert H_{e-r} \vert i \rangle}{(E_{exciton}-E_{\mathrm{pump}}+i\gamma)(E_{exciton}-E_{\mathrm{pump}}+E_{ph}+i\gamma)} \right |^{2}  ,
    \label{eq:Raman}
 \end{equation}
where the numerator holds matrix elements for electron-radiation ($H_{e-r}$) and electron-phonon ($H_{e-ph}$) interactions between the initial ($i$), intermediate ($m$ and $n$), and final ($f$) quantum states. Resonance occurs when the incident ($E_{\mathrm{pump}}$) or scattered ($E_{\mathrm{pump}}-E_{ph}$) photon energy matches the electronic transition energy ($E_{exciton}$). $\gamma$ is a damping factor relating to the inverse lifetime for the resonant scattering process~\cite{Jorio2011}.

\begin{figure}[!htb]
 \centering
 \includegraphics[scale=0.44]{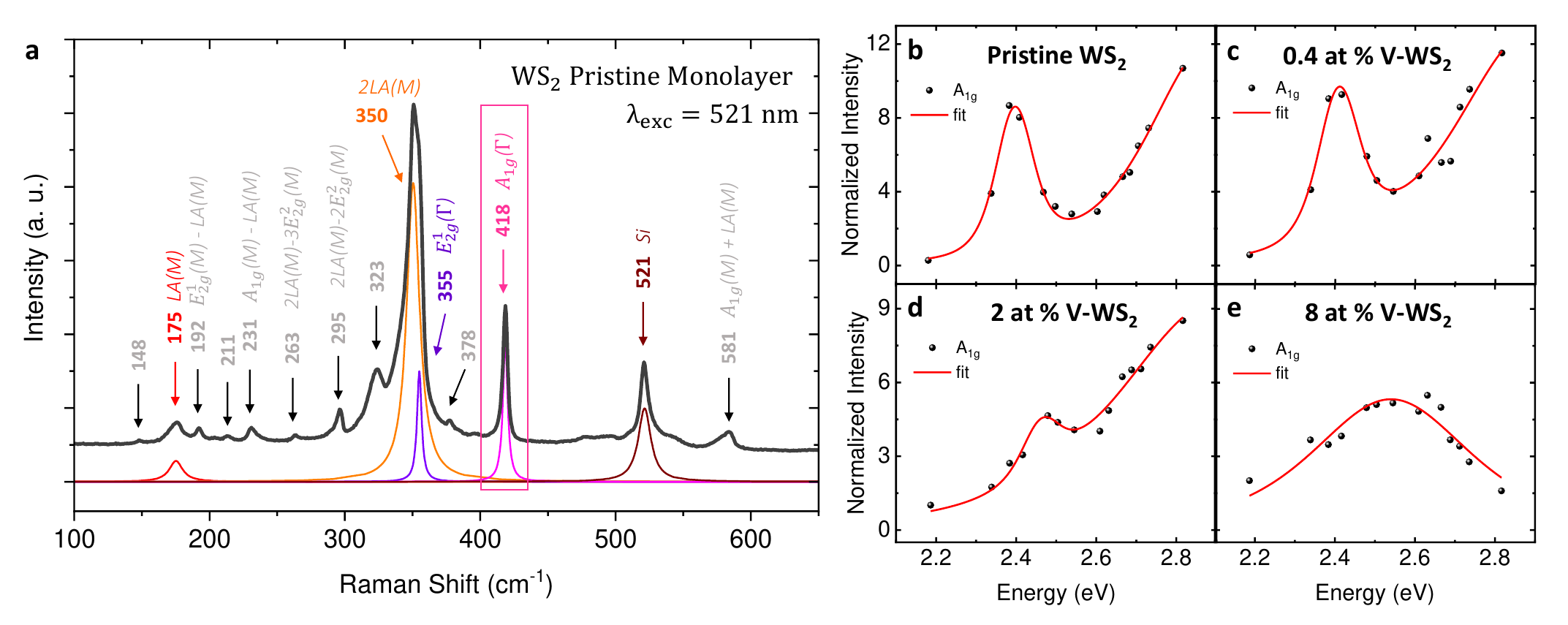} 
 \caption{{\small {\bf a} Raman spectra of a pristine WS$_{2}$ monolayer for 521 nm laser excitation. The peak positions and some of the Raman modes are assigned. 
 The Lorentzian peak fits of selected Raman modes are shown in different colors below the spectra. 
 {\bf b-e} Raman excitation profiles of the $A_{1g}$ Raman mode (highlighted in pink in ({\bf a})) for the ({\bf b}) pristine, ({\bf c}) 0.4 at\%, ({\bf d}) 2 at\%, and ({\bf e}) 8 at\% V-doped WS$_{2}$ monolayers. The solid red curves show the fits to Equation~\ref{eq:Raman}.
  }}
 \label{figRaman}
 \end{figure}

To plot a REP, we need to choose a Raman mode whose intensity can be easily tracked over a wide range of excitation energies. Figure~\ref{figRaman}a shows the Raman spectrum of a pristine \ce{WS2} monolayer with 521 nm excitation. Peak positions and their Raman mode assignments have been investigated by other researchers~\cite{Berkdemir2013,Thripuranthaka2014,McCreary2016-2}. Among these peaks, only the 2LA(M), $E_{2g}$, and $A_{1g}$ modes do not quickly lose discernable signals out of resonance, as shown in Figure~S7. However, 2LA(M) and $E_{2g}$ are not ideal for plotting REP because their energies are almost degenerate, making it difficult to separate their signals, leaving $A_{1g}$ as the best choice for plotting REP.

The $A_{1g}$ intensities for all Raman spectra in Figure S7 (normalized by the Si peak and accounting for its Raman cross section) were fitted by Lorentzian functions and the resulting $A_{1g}$ REP is shown in Figures~\ref{figRaman}b--e. These data points were further fitted by Equation~\ref{eq:Raman}, with the $A_{1g}$ frequency for each sample being the phonon energy $E_{ph}$, and then plotted in red curves. As the pristine, 0.4 at\%, and 2.0 at\% samples present two resonant responses (one centered around 2.4 eV and another at higher energies), we have used two distinct values of $E_{exciton}$ to fit their data. The ascertainment of the resonant energy value around 2.4 eV is not affected by the uncertainty of the value of the higher-energy resonance because of their large spectral distance. The WS$_{2}$ monolayer with 8.0 at\% of vanadium concentration shows only one resonant feature and thus was fitted to a single value of $E_{exciton}$. The lower (B exciton) resonance energies produced by the fittings are $2.42$, $2.44$, $2.49$ and $2.57$ eV for the pristine, 0.4 at\%, 2.0 at\% and 8.0 at\% samples, respectively.  For the pristine sample, the 2.42 eV peak agrees with reported values for the WS$_{2}$ monolayer B exciton energy measured by other groups with the same \cite{DelCorro2016} and different \cite{Zhu2015} techniques. Therefore, it is reasonable to assign the REP peak energy as the B exciton energy in each sample; thus we observe a B exciton blueshift under increasing vanadium concentration. This blueshifting feature is also observed in differential reflectance measurements and in the REP of the $LA$ mode of those samples, as shown in Supporting Figures S6 and S9, respectively.
 
\begin{figure}[!htb]
 \centering
 \includegraphics[scale=0.35]{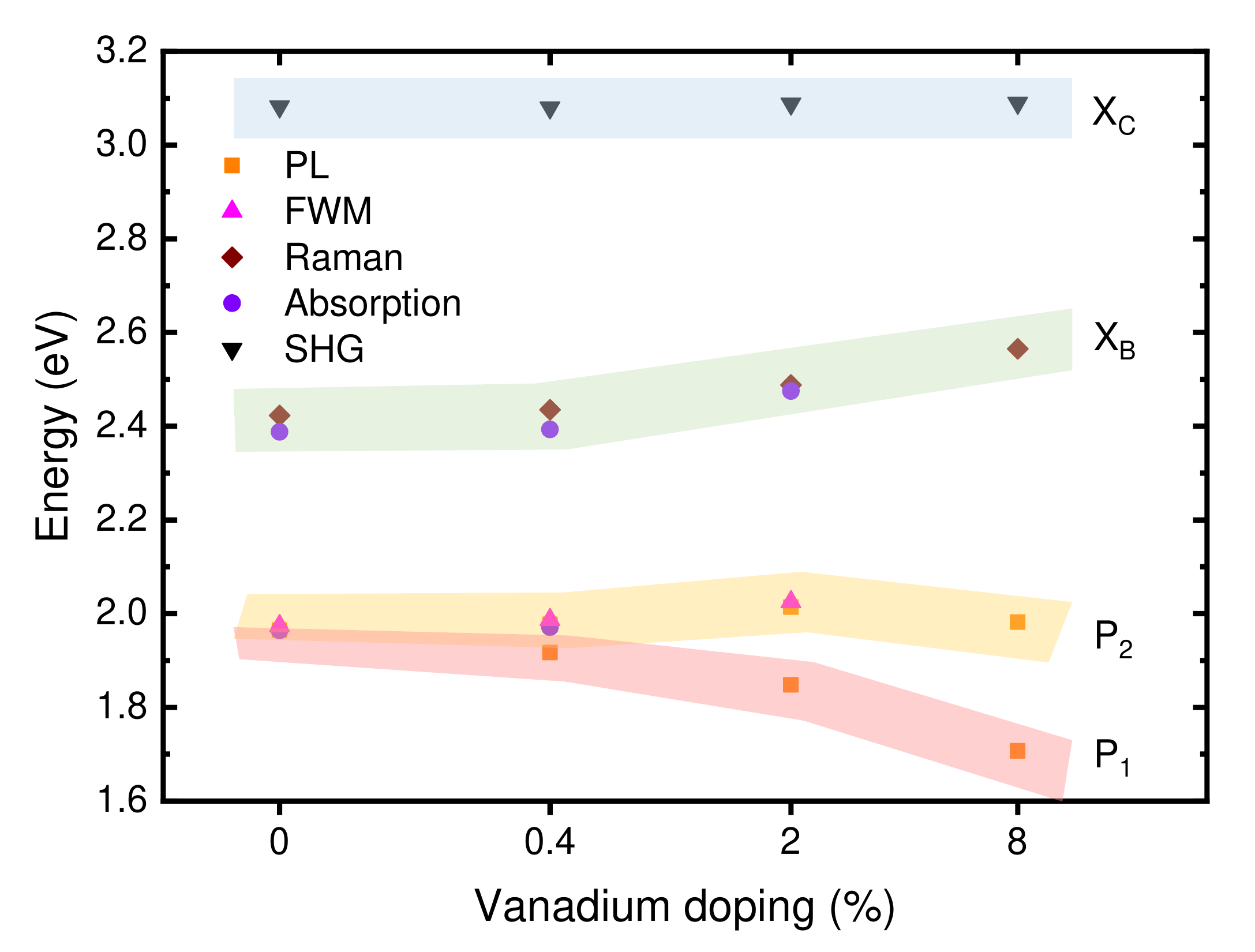} 
 \caption{{\small Summary of P$_{1}$, P$_{2}$, B exciton (X$_{B}$) and C exciton (X$_{C}$) energies for WS$_{2}$ monolayers with respect to their vanadium doping concentration measured by PL, resonant FWM, resonant Raman, absorption (differential reflectance) and resonant SHG measurements. The shadowed areas are guide to the eyes. 
  }}
 \label{fig_Experimental}
 \end{figure}

Resonant SHG was also used to probe the higher-energy transitions in pristine and V-doped WS$_{2}$ samples. As shown in Supporting Figure S10, the pristine sample presents a clear resonance due to the C exciton \cite{Malard2013,Sousa2021}. Furthermore, the V-doped WS$_{2}$ samples (Supporting Figure S10) show a similar resonance energy, indicating that the vanadium doping did not cause any substantial modification to this transition. Hence, beyond the electronic characterization of the samples, the results discussed above also demonstrate an optical method to identify the vanadium doping level in WS$_{2}$ monolayers by measuring the energy splitting and the intensity ratio of P$_{1}$ and P$_{2}$ PL peaks as well as by probing the B exciton energy. To summarize these experimental results, Figure~\ref{fig_Experimental} shows P$_{1}$, P$_{2}$, B exciton and C exciton energies for WS$_{2}$ monolayers with respect to their vanadium concentration for all optical techniques used in this work. In addition, Supporting Figure S11 presents the intensity ratio of P$_{1}$ and P$_{2}$ PL peaks for the pristine and V-doped samples.

\subsection{Electronic structure calculation for V-doped WS$_2$}

\begin{figure}[!htb]
 \centering
 \includegraphics[scale=0.35]{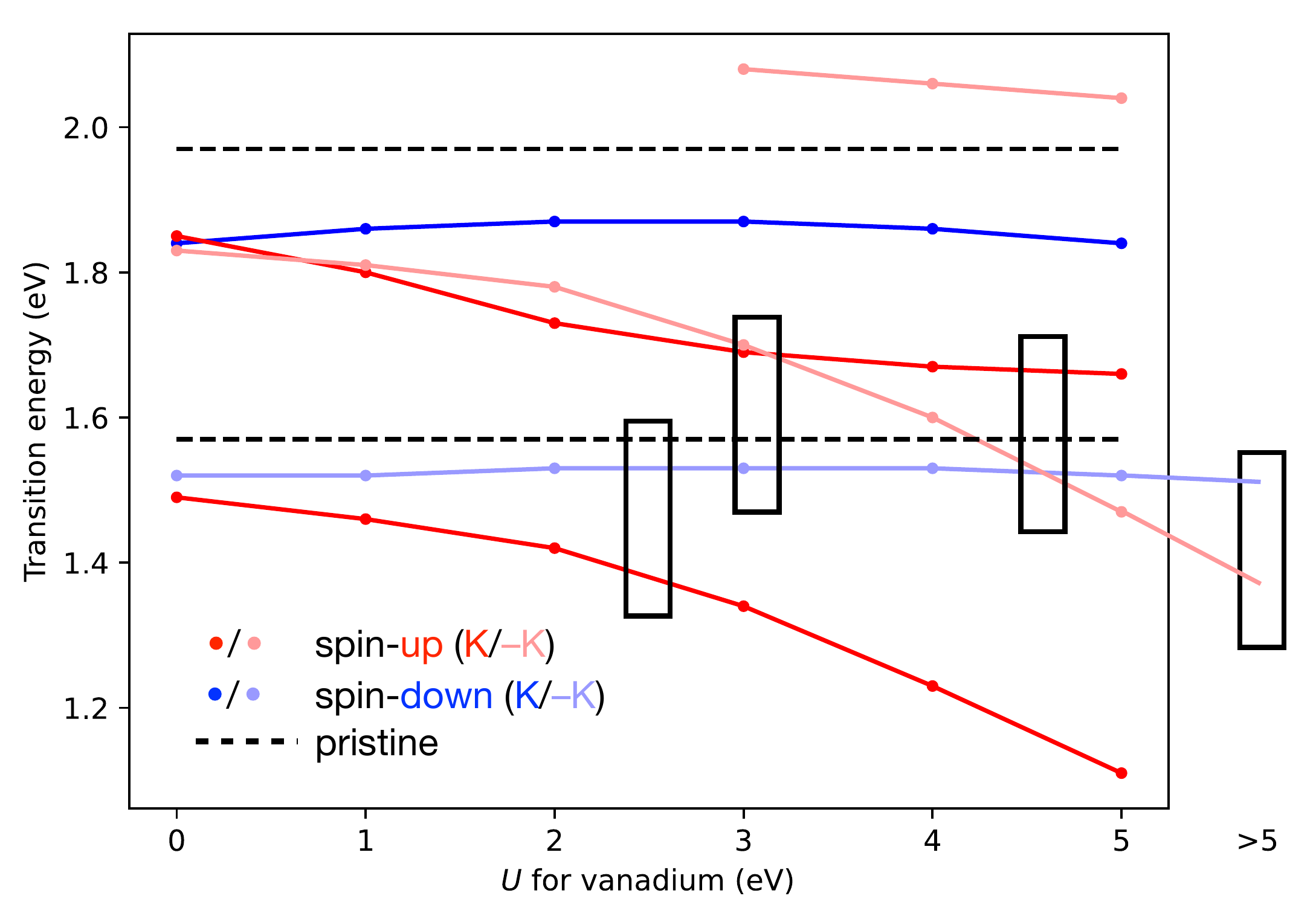}
 \caption{\small Transition energies with largest dipole moments in V-doped \ce{WS2} (1 vanadium in a \(5\times 5\) supercell).
    The dashed line shows the calculated A, B exciton energies in pristine \ce{WS2}.
    The extensions to \(U > \qty{5}{eV}\) are drawn schematically following the trend from the plot.
    Valley degeneracy breaking is non-negligible even without a Hubbard \(U\) correction for vanadium.
    The bandstructure and consequently the optical properties are sensitive to the value of  \(U\).
    As the Hubbard \(U\) increases, the transitions associated with spin-up states redshift while those associated with spin-down states are relatively insensitive to the Hubbard \(U\).
    The rectangles highlight four possible scenarios to explain the \qty{\sim 0.13}{eV} energy difference between the P\textsubscript{1} and  P\textsubscript{2} peaks.
    }
    \label{fig:transition_vs_U}
\end{figure}

\begin{figure}[!htb]
 \centering
 \includegraphics[scale=0.4]{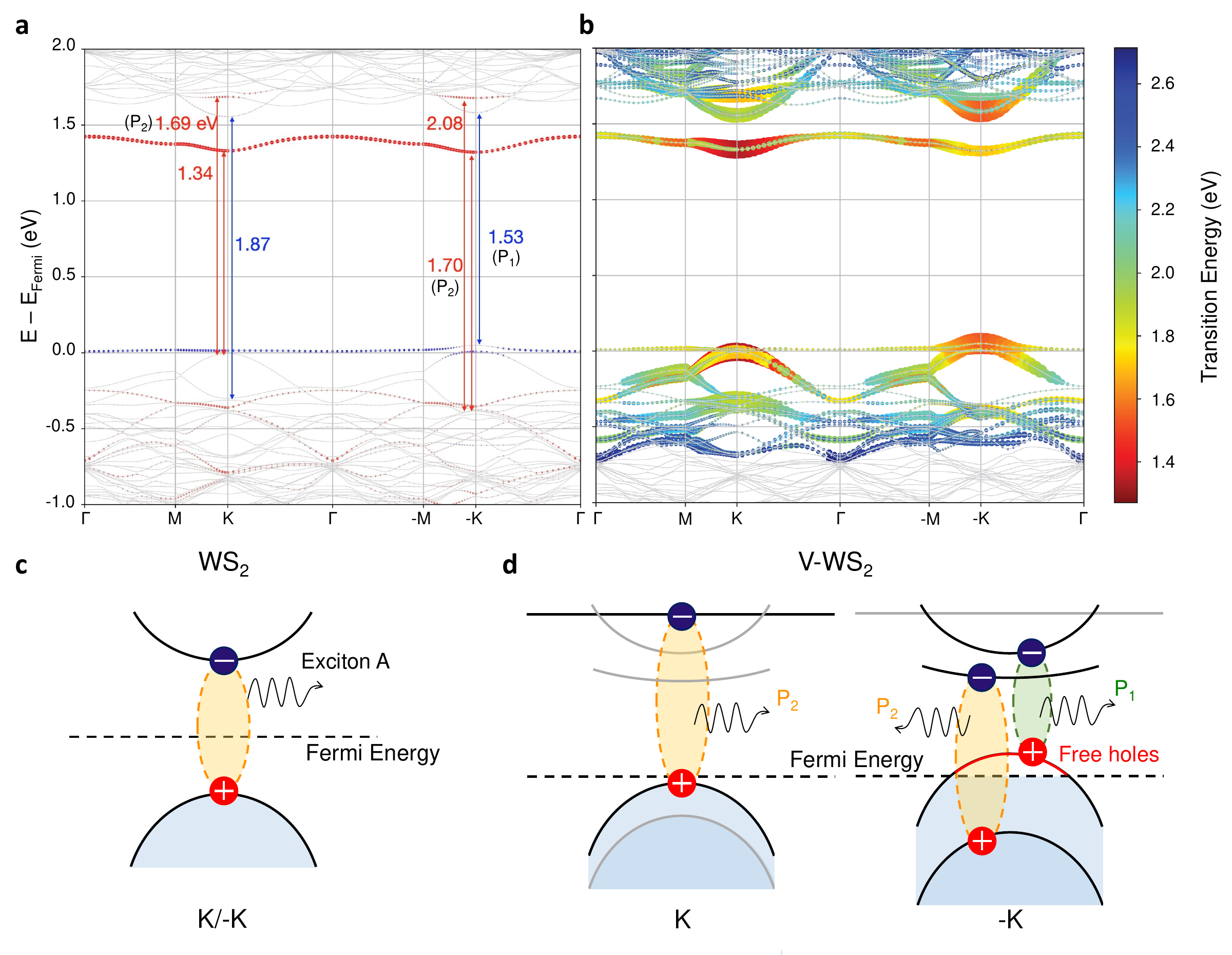}
 \caption{\small {\bf a} Band structure of V-doped \ce{WS2} with \(U = \qty{3.0}{eV}\) for vanadium.
    For each state, a circle is plotted with size proportional to the projection onto the vanadium \(d_{z^2}\) orbital and color indicating the spin (red--up, blue--down).
    The vanadium defect state in the conduction band, mainly with \(d_{z^2}\) character, has spin up  and hybridizes with the spin-up conduction band minimum.
    As the Hubbard \(U\) increases, these two spin-up states decrease in energy, while the spin-down defect states remain far away from the conduction band edge.
    {\bf b} Band structure and transition dipole moments of V-doped \ce{WS2} with \(U = \qty{3.0}{eV}\) for vanadium.
    The energy of transitions at K with the largest transition dipole moments are \qtylist{1.34;1.69;1.87}{eV}.
    The energy of transitions at \(-\)K with the largest transition dipole moments are \qtylist{1.53;1.70;2.08}{eV}.
    {\bf c-d} Schematic illustration of band structure and optical transitions close to the pristine A exciton energy for the pristine and V-doped WS$_{2}$ monolayers with \(U = \qty{3.0}{eV}\) for vanadium. The pristine sample presents its degenerate A exciton recombination at K and \(-\)K valleys ({\bf c}), while the V-doped WS$_{2}$ monolayer displays valley degeneracy breaking with the valence band minima below and above the Fermi level at K and \(-\)K, respectively, showing the presence of free holes at the \(-\)K valley. The black bands are associated with P$_{1}$ or P$_{2}$ optical transitions, while the gray bands are not related to them.
    }
    \label{fig:BS_TDM_U3}
\end{figure}

To understand the optical properties observed in experiment, we examined the band structures (Supporting Figures S12-S17) of V-doped \ce{WS2} with a Hubbard \(U\) correction (\(U = \qtyrange{0}{5}{eV}\)) for vanadium and calculated the transition dipole moments to identify candidate transitions, as summarized in Figure \ref{fig:transition_vs_U}.
The optical transitions in \ce{WS2} mainly relate to the \(d_{x^2}/d_{xy}\) orbitals at the valence band maximum (VBM) and the \(d_{z^2}\) orbitals at the conduction band minima (CBM) at the K or \(-\)K valleys~\cite{liuThreebandTightbindingModel2013}.
The defect states from the vanadium dopant mainly have the character of vanadium \(d_{z^2}\) (see Figure \ref{fig:BS_TDM_U3}a), so their hybridization with the CBM is plausible.
When \(U\) is small (\(U\leq \qty{2}{eV}\)), the defect state in the conduction band is far from the band edge, thus it does not hybridize well with the CBM.
However, the defect state still shows a non-negligible optical coupling to the valence bands at the \(K\) valley.
In this case ($U=\qtylist{0;1;2}{eV}$), the A and B excitons can be easily identified and we find a valley degeneracy breaking of \qtyrange{\sim 0.02}{\sim 0.1}{eV}, A exciton redshifts of \qtyrange{\sim 0.05}{\sim 0.10}{eV}, and B exciton redshifts of \qtyrange{\sim 0.13}{\sim 0.20}{eV} as \(U\) increases.
If \(U\) further increases, the defect state in the conduction band drops in energy and hybridizes with the spin-up CBM (i.e.\ the same spin as the defect state).
The resulting hybridized states both show large optical transition matrix elements with the \(d_{x^2}/d_{xy}\) valence bands.
When \(U\geq \qty{3}{eV}\), we cannot identify the A exciton at the K valley or the B exciton at the \(-\)K valley, so we tentatively call them ``upper'' and ``lower'' spin-up transitions; we will discuss these later.
This hybridization explains why the spin-up transitions are sensitive to the value of \(U\).
On the other hand, the spin-down defect states are far from the CBM and consequently do not hybridize with it, thus the spin-down transitions are insensitive to the value of \(U\).

Identifying the origin of the experimental peaks is not straightforward due to the sensitivity of the band structure to the value of \(U\) in addition to the underestimation of optical transition energies in DFT at the PBE level~\cite{crowleyResolutionBandGap2016}.
We therefore concentrate on the \qty{\sim 0.13}{eV} splitting of P\textsubscript{1} and P\textsubscript{2} around the A exciton energy region in \cref{figPL} and find 4 possible scenarios with \(U \qtylist{\sim 2.5;\sim 3;\sim 4.5;>5}{eV}\), as highlighted in the inset black rectangles of Figure \ref{fig:transition_vs_U}.
In the \(U\qty{\sim 2.5}{eV}\) scenario, the P\textsubscript{1}--P\textsubscript{2} energy difference is from the valley degeneracy breaking of the A exciton.
In the \(U\qty{\sim 3}{eV}\) scenario, P\textsubscript{1} is the A exciton at \(-\)K and P\textsubscript{2} could be the upper spin-up transition at K and/or the lower transition at \(-\)K since their energies coincide.
In the \(U \qty{\sim 4.5}{eV}\) scenario, the A exciton and the lower transition energies at \(-\)K have similar energies, so they both could explain the P\textsubscript{1} peak, while the upper spin-up transition at K could be the P\textsubscript{2} peak.
In the \(U\qty{>5}{eV}\) scenario, P\textsubscript{1} is the lower spin-up transition and P\textsubscript{2} is the A exciton at \(-\)K.
Among these scenarios, \(U\qtylist{\sim 3; \sim 4.5}{eV}\) can qualitatively explain the blueshift of the P\textsubscript{2} peak relative to the pristine A exciton. 
In the \(U\qtylist{\sim 3; \sim 4.5; >5}{eV}\) scenarios, the high energy of the upper spin-up transition could qualitatively explain the blueshift of the B exciton while the transition with the lowest energy (the lower spin-up transition at K) may lead to PL emission with further lower energy, however we could not observe this emission given our instrument sensitivity.
Moreover, the \(U\qtylist{\sim 3}{eV}\) scenario shows occupied/unoccupied VBM at K/$-$K, that is in agreement with our power-dependent PL, resonant FWM, and differential reflectance measurements in which radiative recombinations to unoccupied/occupied bands for P$_{1}$/P$_{2}$ PL peaks are shown (the band structure representation of these recombinations is schematically depicted in Figures \ref{fig:BS_TDM_U3}c-d).
The linear power dependence of P\textsubscript{2} could also be explained since the electron state of the P\textsubscript{2} at $-$K shows non-negligible dispersion, meaning this state has a significant contribution from the \ce{WS2} basal plane which can support the linear power dependence.
Although none of these scenarios can fully explain all of the optical responses observed in the experiments, they provide relevant insights into the electronic structure modifications by the introduction of vanadium on a WS$_2$ monolayer as well as its dependence on the Hubbard \(U\) correction.

\section{Conclusion}
 
In summary, by applying broadband optical characterization through the use of several optical tools we studied the electronic band structure and optical transitions, as well as their dependence on vanadium doping, of \ce{WS2} grown by single-step CVD.
Power-dependent PL measurements showed that the A exciton PL peak of the pristine WS$_{2}$ monolayer split into two PL peaks after vanadium doping --- one (P$_1$) redshifted and another (P$_2$) blueshifted with respect to the pristine A exciton PL peak.
The linear/sublinear power dependence of P$_2$/P$_1$, together with the fact that FWM resonant profiles only show the P$_2$ peak, indicated that the P$_2$/P$_1$ PL peak is associated with a radiative recombination to an occupied/unoccupied electronic band.
Resonant Raman spectroscopy measurements revealed a blueshift in the B exciton energy under increasing vanadium doping, while the SHG resonant profile presented no modifications in the C exciton energy.
First-principle calculations showed valley degeneracy breaking after vanadium doping and a strong dependence of the band structure on the Hubbard $U$ parameter for vanadium.
Both our experimental and computational results indicated the presence of free holes in V-doped \ce{WS2}, suggesting a similar magnetic exchange mechanism as in the case of V-doped \ce{WSe2}.\cite{Duong2019,Song2021} 
Our work shows the great potential of broadband optical characterization to study the impact of defect in 2D materials.

\section{Methods}

\subsection{Sample Preparation}

In the liquid-assisted CVD method used to synthesize the monolayer V-doped WS$_{2}$ flakes, ammonium metatungstate hydrate ((NH$_{4}$)$_{6}$H$_{2}$W$_{12}$O$_{40}$·xH$_{2}$O) and sodium cholate hydrate (C$_{24}$H$_{39}$NaO$_{5}$·xH$_{2}$O) powders were first mixed and dissolved in DI water to form the W precursor solution. A vanadyl sulfate (VO[SO$_{4}$]) powder was separately dissolved in DI water to form the V precursor solution. The two solutions were mixed and spin-coated onto SiO$_{2}$/Si substrates, and the as-treated substrates were placed in a quartz tube as the reaction chamber. A sulfur powder was placed upstream in the quartz tube, and with ultrahigh purity Ar carrier gas streaming, the quartz tube was placed in a furnace and heated to 825 °C for 15 min for the precursors to react. After the reaction, the furnace was cooled to room temperature naturally and with Ar protection. Different dopant concentrations were realized by tuning the volume ratio between the W and V precursor solutions.

\subsection{Spectroscopy Measurements}

The FWM and SHG measurements were performed by using a picosecond OPO laser (\textit{APE} Picoemerald), with a pump power of 5 mW for both $\omega_{\mathrm{pump}}$ and $\omega_{1064}$ for the FMW and 3 mW for the SHG. The power-dependent PL measurements were performed using a 561 nm CW diode laser. The laser beam was focused on the sample by a 40$\times$ objective with numerical aperture N.A. = 0.95. The backscattered signal was collected by the same objective and directed to the spectrometer (\textit{Andor} Shamrock 303i) equipped with a sensitive CCD camera (\textit{Andor} IDUS DU401A-BV). Different sets of filters were used to reject laser light from the spectrometer: 750 nm or a 842 nm shortpass filter for SHG and FWM and 561 nm longpass filter for PL measurements. The differential reflectance measurements were performed with a white light source focused on the sample by the same objective and directed to the same spectrometer.

Resonant Raman spectroscopy measurements were performed on a HORIBA \textit{Jobin Yvon} T64000 triple-monochromator spectrometer equipped with a CCD detector and using a 1800 g/mm diffraction grating. The samples were excited by a Ar-Kr laser to cover excitation energies from 2.18 to 2.81 eV and with an incident power of 0.6 mW. A 100$\times$ objective with numerical aperture N.A. = 0.9 was used to focus the laser beam and collect the backscattered signal.

\subsection{Calculations}
Density functional theory (DFT) is carried out by VASP~\cite{kresseEfficiencyAbinitioTotal1996,kresseEfficientIterativeSchemes1996,kresseInitioMolecularDynamics1993,kresseUltrasoftPseudopotentialsProjector1999} at the PBE level~\cite{perdewGeneralizedGradientApproximation1996} with cutoff energy 500 eV and a \(5\times 5\times 1\) Monkhorst-Pack mesh~\cite{monkhorstSpecialPointsBrillouinzone1976}.
The lattice constant for the supercell (1 vanadium doped in a \(5\times 5\) \ce{WS2}) is kept 5 times the lattice constant of primitive \ce{WS2} (\(5\times \qty{3.188}{\AA} = \qty{15.94}{\AA}\)), assuming that the low doping level would have little effect on the lattice constant.
The convergence criteria for electronic self-consistency and force relaxation are \qty{1e-7}{eV} and \qty{0.01}{eV/\AA} on all atoms.
Spin-orbit coupling is included after the relaxation process for electronic structure calculations.
The rotationally invariant DFT\(+U\) method~\cite{liechtensteinDensityfunctionalTheoryStrong1995} is used for the Hubbard \(U\) correction.
Transition dipole moments are calculated with the code \texttt{VaspBandUnfolding}~\cite{zhengVaspBandUnfolding2023}.

\section*{Author Contributions}

F.B.S., B.Z., M.T., V.H.C. and L.M.M. conceived the ideas. F.B.S. and L.M.M. performed PL, FWM, SHG and differential reflectance measurements. F.B.S., G.C.R., M.A.P. and L.M.M. carried out resonant Raman experiments. F.B.S. and L.M.M. analyzed the experimental data. M.L., D.Z and M.T. prepared the samples and performed the STEM experiments. B.Z. and V.H.C. performed the DFT calculations and analyzed the theoretical data. F.B.S, B.Z., V.H.C. and L.M.M wrote the manuscript. All authors reviewed and agreed on the final version of the manuscript.

\begin{acknowledgement}

F.B.S, G.C.R, M.A.P. and L.M.M acknowledge financial support from CNPq, CAPES, FAPEMIG, FINEP, Brazilian Institute of Science and Technology (INCT) in Carbon Nanomaterials and Rede Mineira de Materiais 2D (FAPEMIG). 
B.Z. and V.H.C. acknowledge support from the 2DCC-MIP which is funded by NSF cooperative agreement DMR-2039351. 
M.T. and D.Z. also thank AFOSR for financial support (FA9550-23-1-0447).

\end{acknowledgement}






\end{document}